\documentclass[twocolumn,prl,superscriptaddress,showpacs,byrevtex]{revtex4}
\usepackage{epsf,graphicx,amssymb}
\usepackage[usenames]{color}
\usepackage{natbib}
\usepackage{float}
\usepackage{gensymb}
\usepackage{amsmath,amssymb}
\usepackage{color}
\begin{document}

\title{Wave turbulence on the surface of a fluid in a high-gravity environment}

\author{A. Cazaubiel}
\affiliation{Universit\'e de Paris, Univ Paris Diderot, MSC, UMR 7057 CNRS, F-75 013 Paris, France}
\author{S. Mawet}
\affiliation{GRASP, D\'epartement de Physique B5a, Universit\'e de Li{\`e}ge, FNRS - B-4000 Li{\`e}ge, Belgium}
\author{A. Darras}
\affiliation{GRASP, D\'epartement de Physique B5a, Universit\'e de Li{\`e}ge, FNRS - B-4000 Li{\`e}ge, Belgium}
\affiliation{Experimental Physics, Saarland University, D-66123 Saarbr{\"u}cken, Germany}
\author{G. Grosjean}
\affiliation{GRASP, D\'epartement de Physique B5a, Universit\'e de Li{\`e}ge, FNRS - B-4000 Li{\`e}ge, Belgium}
\author{J. J. W. A. van Loon}
\affiliation{Gravity Simulation Laboratory, ESTEC, ESA, Noordwijk, The Netherlands}
\affiliation{ACTA, VU University, Amsterdam, The Netherlands}
\author{S. Dorbolo}
\affiliation{GRASP, D\'epartement de Physique B5a, Universit\'e de Li{\`e}ge, FNRS - B-4000 Li{\`e}ge, Belgium}
\author{E. Falcon}
\email[E-mail: ]{eric.falcon@univ-paris-diderot.fr}
\affiliation{Universit\'e de Paris, Univ Paris Diderot, MSC, UMR 7057 CNRS, F-75 013 Paris, France}
\date{\today}
\begin{abstract}  
We report on the observation of gravity-capillary wave turbulence on the surface of a fluid in a high-gravity environment. By using a large-diameter centrifuge, the effective gravity acceleration is tuned up to 20 times the Earth gravity. The transition frequency between the gravity and capillary regimes is thus increased up to one decade as predicted theoretically. A frequency power-law wave spectrum is observed in each regime and is found to be independent of the gravity level and of the wave steepness. While the timescale separation required by weak turbulence is well verified experimentally regardless of the gravity level, the nonlinear and dissipation timescales are found to be independent of the scale, as a result of the finite size effects of the system (large-scale container modes) that are not taken currently into account theoretically.


\end{abstract}


\maketitle
\paragraph*{Introduction. \textemdash}Wave turbulence concerns the study of random nonlinear waves in interaction. Although this phenomenon occurs in various situations (ocean surface waves, plasma waves, hydro-elastic or elastic waves, internal waves, or gravitational waves), it is far from being completely understood \cite{Falcon2010revue,Zakharovbook,Nazarenkobook,Newell2011,Shrira2013,Hawai}. For instance, most laboratory experiments on gravity wave turbulence are not in agreement with weak turbulence theory \cite{FalconPRL07,DenissenkoPRL07,CobelliPRL11,DeikeJFM15,AubourgPRF17}. Several reasons are given such as the roles of the dissipation, of the strong nonlinear waves, or of the basin finite size that are usually not taken into account theoretically. The goal here is to better study gravity wave turbulence by tuning the natural key parameter seldom modified until now, the gravity field.

In this Letter we describe a unique experimental setup that is designed to build and study hydrodynamic wave turbulence on the surface of a fluid in hypergravity conditions in a laboratory.  The relevant parameters of wave turbulence depend strongly on the gravity acceleration $g^*$ at different levels. First, it appears in the dispersion relation of linear waves $\omega(k,g^*)=\sqrt{g^*k + \frac{\gamma}{\rho} k^3}$
(with $\omega=2\pi f$ the wave angular frequency, $k$ its wavenumber, $\rho$ the fluid density and $\gamma$ its surface tension). Second, it appears in the transition frequency between the capillary and gravity wave turbulence regimes \cite{FalconPRL07}
\begin{equation}
f_{gc}=\frac{1}{\sqrt{2}\pi} \left({\frac{\rho}{\gamma}}\right)^{1/4}{g^*}^{3/4} \ {\rm .}
\label{fgctheo}
\end{equation}
Third, the timescale of nonlinear interactions between gravity waves, $\tau^g_{nl}\sim \varepsilon^{-2/3}k^{-3/2}{{g^*}^{1/2}}$, is assumed slow compared to the linear wave period $\tau_{lin}=1/f$, to require a weak nonlinearity $\sim \tau_{lin}/\tau^g_{nl} \sim \varepsilon^{2/3}k^{1}/g^* \ll 1$ ($\varepsilon$ being the energy flux) \cite{Newell2011}. As nonlinearity increases with the scale $k$, breaking of weak turbulence  for gravity waves is expected to occur at small scale ($k^g_c \sim g^*/\varepsilon^{2/3}$), and even smaller with higher $g^*$. Another relevant theoretical parameter is the critical energy flux $\varepsilon_{c}=(\gamma g^*/\rho)^{3/4}$ breaking the weak turbulence at the transition scale  \cite{NewellPRL92}.  Such dependences of these parameters on $g^*$ are in favor of the study of wave turbulence in a high-gravity level.  

Tuning the gravity level, e.g. from $1$ to $20$ times the Earth acceleration $g$, will thus increase $f_{gc}$ and $\varepsilon_{c}$ up to a factor 10; these two parameters being hardly modified differently since $(\rho / \gamma)$ is roughly constant for most of usual fluids. This is also an astute solution, not reported so far, to expand significantly in laboratory the inertial range of observation of gravity wave turbulence. The latter is usually limited (1 decade at most) at large scales by the forcing ($\sim$ Hz due to wavemaker limitations) and at small scales by the capillary regime near $f_{gc}$ ($\simeq $ the ten of Hz for most of fluids at $1g$). Tuning the gravity homogeneously is nevertheless far from being simple. Indeed, fastly rotating a liquid in a small-size container generates a high centrifugal acceleration but with a radial gradient of effective gravity leading to a parabolic curvature of the liquid surface. To circumvent this effect, we use a large-scale centrifuge with a free swinging gondola that generates an effective gravity still perpendicular to the liquid surface at rest. Note that tuning to low-$g$ values was performed to observe pure capillary wave turbulence on more than 2 decades with no effect of gravity waves during parabolic flights ($10^{-2}g$) \cite{FalconEPL09} or on-board of the International Space Station ($10^{-5}g$) \cite{BerhanuEPL19}. 

Here, we report the observation of surface wave turbulence in hypergravity, $f_{gc}$ being increased up to one decade. We also show that finite size effects modify the usual energy transfer mechanism (nonlinear wave interactions); finite size effects having rarely been addressed experimentally \cite{DeikeJFM15,HassainiPRF18} compared to numerical or theoretical studies \cite{ZakharovJETP05,NazarenkoJSM06,LvovPRE10,PanJFM17}.



 
\begin{figure}[!h]
\begin{center}
\includegraphics[width=75mm]{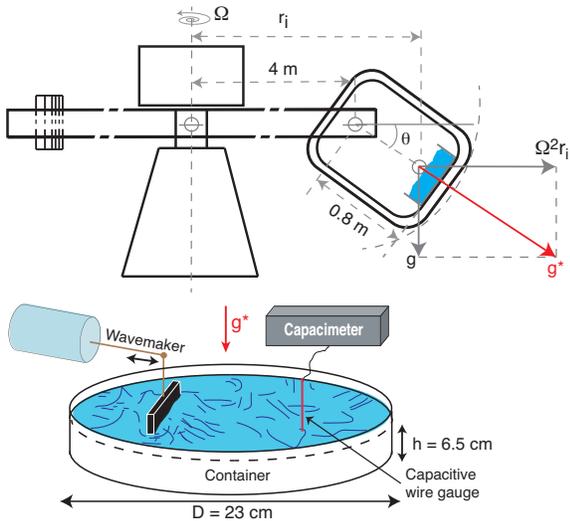}
\caption{Experimental setup. Top: Schema of the large diameter centrifuge showing the generation of an apparent gravity normal to the fluid surface. Bottom: Experimental setup inside the gondola. See movies in Supplemental Material \cite{SuppMat}.}
\label{fig01}
\end{center}   
\end{figure}

\paragraph*{Experimental setup.\textemdash}Experiments are performed in the ESA Large Diameter Centrifuge (LDC), an 8 m diameter four-arm centrifuge with free-swinging gondola on each arm (see top of Fig. \ref{fig01}). The gondola containing the experiment is locked at the end of an arm, and a counter-weight is placed on the other end. The maximum angular velocity $\Omega$ of the centrifuge is 67 rpm (i.e. 101 km/h at the arm's end, 4 m from the axis). Due to centrifugal acceleration, the apparent gravity is $\vec{g}^{\, *}=\vec{g}+ \Omega^2 r_i \vec{e}_r$, with $g=9.81$ m/s$^{2}$ the Earth gravitational acceleration, $r_{i}$ the radius to the point of interest. The range of effective gravity $g^*=\sqrt{g^2+(\Omega^2 r_i)^2}$ is 1 to 20 times Earth gravity. The angle $\theta=\arctan{(\Omega^2 r_i/g)}$ of the gondola in rotation with the horizontal decreases rapidly from $90^{\circ}$ when $\Omega$ increases and is less than $12^{\circ}$ when $g^*>5g$. The experimental setup placed in the gondola bottom consists of a cylindrical container, 23 cm in diameter, filled with distilled water (density $\rho=1000$ kg/m$^3$ and surface tension $\gamma=74$ mN/m) up to a depth $h=6.5$ cm to reach a deep water regime ($\lambda \ll 2\pi h$).  Surface waves are generated by a rectangular flap wavemaker driven by a servo-motor moving according to a random noise forcing of narrow bandwidth (3 to 5 Hz) and of rms amplitude, $\sigma_A <  2.2$ cm. The wave height $\eta(t)$ is monitored at a given location by a homemade capacitive wire gauge during $\mathcal{T}=1000$ s with a 10 $\mu$m sensitivity \cite{FalconPRL07}. Typical wave steepness is less than 0.1. The fluid is filled in the container once the prescribed acceleration is reached to avoid a possible fluid rotation, as controlled by an on-board camera in the gondola. During stationary rotation but with no wave forcing, the free surface of the fluid remains perfectly parallel to the container bottom, as expected. No Coriolis force effect is observed on the wave field since the wave velocities are much smaller than the centrifuge velocity. 

\begin{figure}[!t]
\begin{center}
\includegraphics[scale=0.45]{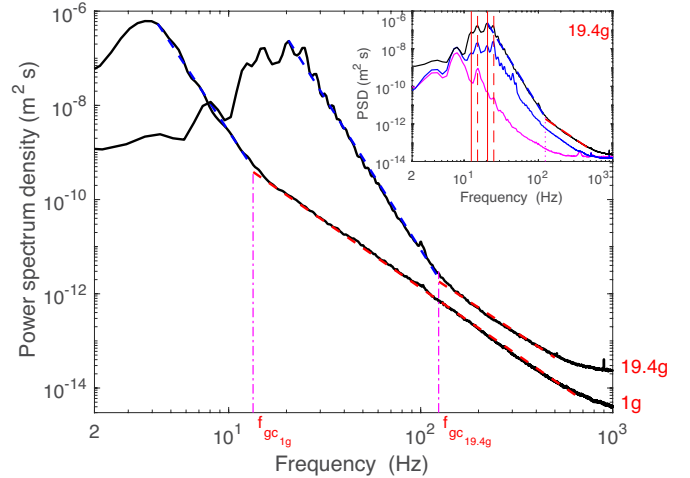}
\caption{Power spectrum density (PSD) of the wave height $\eta(t)$ for $1g$ and $19.4g$. Dashed lines display best power-law fits for (blue) gravity and  (red) capillary regimes. Vertical dash-dotted lines correspond to $f_{gc}(g^*)$ of Eq.\ (\ref{fgctheo}). $\sigma_A=15.5$ mm. Inset: PSD for different vibration amplitudes $\sigma_A=0$, 3.7 and 15.5 mm (bottom to top). $g^*=19.4g$. Solid (resp. dashed) lines are the first axisymmetrical $m=0$ (resp. non-axisymmetrical $m=1$) basin modes solutions of $J'_m(k_nD/2)=0$,  with $J'_m(\cdot)$ the derivative of the $m$-th order Bessel function of the first kind \cite{Lamb}. }  
\label{fig02}
\end{center}
\end{figure}

\paragraph*{Wave spectrum.\textemdash}For a fixed $g^*$, the temporal evolution of the wave height $\eta(t)$ is erratic and its rms value, $\sigma_{\eta}=\sqrt{\langle \eta \rangle^2_t}$ of few mm, increases linearly with the forcing amplitude $\sigma_A$ regardless of $g^*$. The spectrum of $\eta(t)$ is shown in Fig.\ \ref{fig02} for two different apparent gravities, 1$g$ and close to 20$g$. At 1$g$, we observe two frequency power-law regimes corresponding to the gravity and capillary wave turbulence regimes as already reported previously \cite{FalconPRL07}. For a gravity close to 20$g$, the power law regimes are still observed and the transition frequency $f_{gc}$ between both regimes increases significantly of 1 decade as expected (see below). Consequently, the inertial range of the gravity regime has been also extended to reach roughly 1 decade. Nevertheless, note that the maximum of the spectrum occurs at a higher frequency at high-$g$ level. This comes from the gravity dependence of the dispersion relation. Indeed, the forcing frequency bandwidth at 1$g$ corresponds to wavelengths $\lambda$ less than the container diameter $D$, whereas at high $g^*$ it corresponds to $\lambda > D$, the maximum of the spectrum being then related to the first eigenmodes of a cylindrical basin (notably $\lambda \sim D/3$), as shown in the inset of Fig.\ \ref{fig02}.

\begin{figure}[t]
\begin{center}
\includegraphics[scale=0.45]{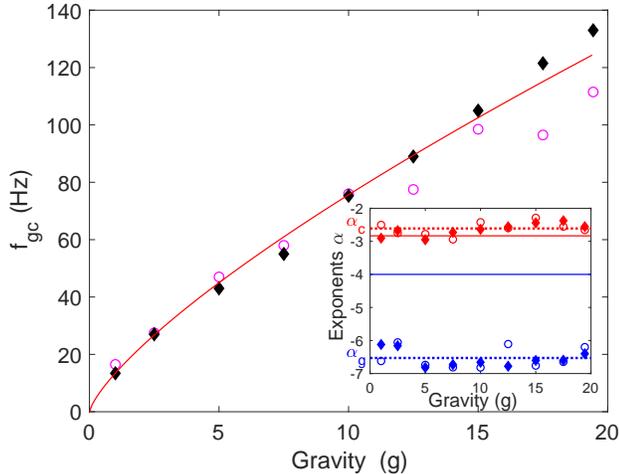} 
\caption{Transition frequency $f_{gc}$ between gravity and capillary wave turbulence regimes versus $g^*$. Vibration amplitude: $\sigma_A=$ ($\circ$) 3.7, ($\blacklozenge$) 15.5 mm. Solid line is the prediction of Eq.\ (\ref{fgctheo}). Inset: Frequency power-law exponents $\alpha$ of the wave spectrum for capillary (red) and gravity (blue) regimes. Dotted lines: mean values $\langle \alpha_c\rangle_{g^*}$ and $\langle \alpha_g\rangle_{g^*}$. Solid lines: weak turbulence predictions (see text).} 
\label{fig03}
\end{center}
\end{figure}

\paragraph*{Transition frequency.\textemdash}The transition frequency $f_{gc}$ between gravity and capillary wave turbulence regimes is inferred from the wave spectrum, and is plotted in Fig. \ref{fig03} versus $g^*$ for two forcing amplitudes. $f_{gc}$ is found to increase strongly with $g^*$ by 1 decade in very good agreement with the prediction of Eq.\ (\ref{fgctheo}) with no fitting parameter. The frequency power-law exponents of the spectrum in both regimes are reported in the inset of Fig.\ \ref{fig03} as a function of $g^*$. They are found to be independent of $g^*$ and of the wave steepness within the range of parameters. In the capillary regime, the experimental exponent value $\alpha_c \simeq -2.6$ (red dotted line) is in good agreement with the prediction $-17/6 \simeq -2.8$  (red solid line) of the weak turbulence spectrum $S_{\eta}^c \sim \varepsilon^{1/2}\left(\frac{\gamma}{\rho}\right)^{1/6}f^{-17/6}$ \cite{Zakharov67Cap}. In the gravity regime, one has $\alpha_g \simeq -6.4$ (blue dotted line) that differs significantly from weak turbulence prediction $S_{\eta}^g \sim \varepsilon^{1/3}gf^{-4}$ (blue solid line) \cite{Zakharov67Grav} as reported earlier at $1g$ in different basin sizes  (0.5 - 50 m) \cite{FalconPRL07,DenissenkoPRL07,CobelliPRL11,DeikeJFM15,AubourgPRF17}. This departure has been recently shown to be related to the modulation of coherent nonlinear structures (bound waves) \cite{MichelPRF18}. We also compute the probability density functions of $\eta(t)$. They are found to be independent of $g^*$ for the same forcing amplitude, and to be roughly described by a Tayfun distribution (the first nonlinear correction to a Gaussian) as a confirmation of weak nonlinearity of the wave field (see Supplemental Material). 


\begin{figure}[t]
\begin{center}
\includegraphics[scale=0.45]{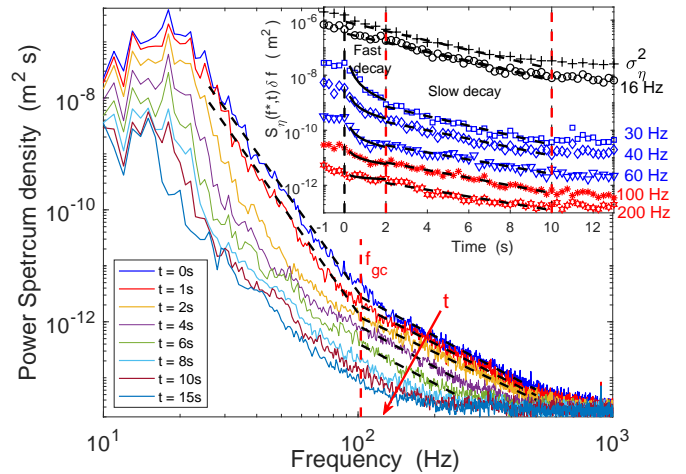} 
\caption{Temporal decay of the wave spectrum $S_{\eta}(f,t)$ for $15g$. Inset: Semilogy-plot of the decay of wave energy $\sigma^2_{\eta}$ ($+$), and of the Fourier modes $S_{\eta}(f^{\ast},t)\delta f$ at different frequencies $f^{\ast}$ corresponding either to the main container mode ($\circ$), the gravity regime (blue symbols) or the capillary regime (red symbols). Dashed vertical lines delimits the fast power-law decay (solid-line fits) and then the slow exponential decay (dashed-line fits at different times). $\sigma_A=15.5$ mm. } 
\label{fig04}
\end{center}
\end{figure}

\paragraph*{Decay experiments.\textemdash}We now present nonstationnary experiments to obtain estimations of the nonlinear and dissipation timescales of wave turbulence, and of the mean energy flux cascading through the scales. Once a stationary wave turbulence is reached, we stop the wavemaker at time $t=0$, and the temporal decay of the wave height is recorded during $\mathcal{T}=15$ s. The experiment is automatically iterated up to 330 times to improve statistics and results are averaged. 

For $g^*=15g$, the temporal decay of the spectrum is shown in Fig.~\ref{fig04}, the spectrum being computed over short-time intervals $\delta t$. The power laws are found to be conserved at the very beginning of the decay in the gravity regime and over a much longer duration in the capillary regime, before vanishing in favor of a purely dissipative regime. When iterating this experiment for different $g^*$, we found that this self-similar decay in the first moments is independent of the gravity level (see Supplemental Material).  This self-similar decay of the wave spectrum has been predicted theoretically \cite{Zakharovbook} and observed at $1g$ either in the gravity regime \cite{DeikeJFM15} or in the capillary regime \cite{DeikePRE12}. 

The temporal evolution of the amplitude of the Fourier modes inferred from the decaying spectrum is then shown in the inset of Fig.\ \ref{fig04}. First, a fast time power-law decay is observed ($t \leq 2$ s) due to nonlinear mechanisms but that differs from $t^{-1/2}$ (predicted for pure 4-wave interactions) in the gravity regime \cite{Bedard2013} and from $t^{-1}$ (for pure 3-wave interactions) in the capillary regime \cite{FalkovichEPL95} (see discussion below). Then, a slow exponential decay is observed ($2 \leq t \leq 10$ s) due to a viscous linear damping of a large-scale mode. Indeed, each Fourier mode is found to decay exponentially with the same rate given by the viscous damping of the main container mode $\omega(k_{D/3})$ (see $\circ$-symbols). When compared to the total wave energy decay [computed from the wave variance decay $\sigma^2_{\eta}$ ($+$-symbols)], most of energy is contained at this large scale mode and plays the role of an energy source which sustains the turbulent cascade.  We emphasize that this dynamics is different from the decay of non-interacting waves where high frequencies are usually damped faster by linear viscosity than low ones. Here, it is due both to cumulative energy transfer from the large-scale mode and to transfer by nonlinear interactions.

\begin{figure}[t]
\begin{center}
\includegraphics[scale=0.45]{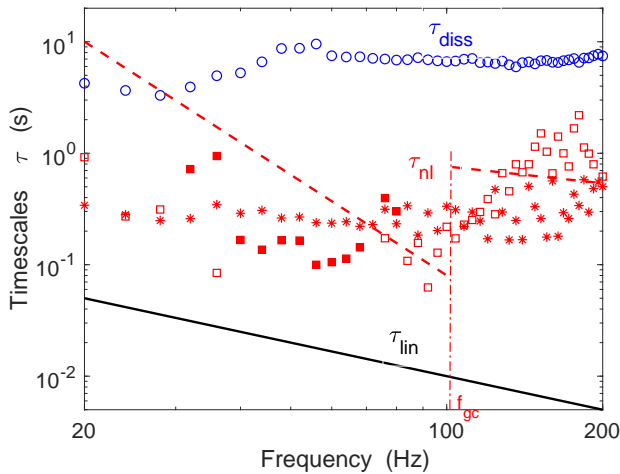} 
\caption{Wave turbulence timescale separation for $g^*=15g$. (Solid line) Linear timescale $\tau_{\mathrm{lin}}=1/f$. Nonlinear timescale $\tau_{\mathrm{nl}}$ estimated from the fast decay power-law fit of Fig.\ \ref{fig04} ($*$) or from the phenomenological model of Eq. (\ref{decaymodel}) $1/b$ ($\blacksquare$) and $1/c$ ($\square$). Dashed red lines are predictions of $\tau_{\mathrm{nl}}$ in the capillary and gravity regimes with constants chosen to $c_g=c_c=1$.  ($\circ$) Dissipation timescale, $\tau_{\mathrm{diss}}$, estimated from the slow decay of Fig.\ \ref{fig04}. $\sigma_A=15.5$ mm.} 
\label{fig05}
\end{center}
\end{figure}

\paragraph*{Timescale separation.\textemdash}Let us now verify the weak turbulence theory assumption of a timescale separation $\tau_{\mathrm{lin}}(f) \ll \tau_{\mathrm{nl}}(f) \ll  \tau_{\mathrm{diss}}(f)$ between the linear propagation time $\tau_{\mathrm{lin}}=1/f$, the nonlinear interaction time $\tau_{\mathrm{nl}}$, and the dissipation time $\tau_{\mathrm{diss}}$. $\tau_{\mathrm{nl}}(f)$ is obtained by a power-law fit of the fast decay on 1 decade in time (0.2 to 2 s) of each Fourier mode, and $\tau_{\mathrm{diss}}(f)$ by an exponential fit of the slow decay.  For a fixed $g^*$, the timescale separation is well verified as shown in Fig.\ \ref{fig05}.  More surprisingly, $\tau_{\mathrm{nl}}$ ($*$) and $\tau_{\mathrm{diss}}$ ($\circ$) are found to be roughly independent of the scale $f$, contrary to weak turbulence predictions in the gravity ($\tau^g_{\mathrm{nl}}=c_g \varepsilon^{-2/3}g^{2}f^{-3}$) and capillary [$\tau^c_{\mathrm{nl}}=c_c\varepsilon^{-1/2} (\rho/\gamma)^{-1/2}f^{-1/2}$] regimes (see dashed lines in Fig. \ref{fig05}, $\varepsilon$ being estimated experimentally as below). This departure is ascribed to the finite size effects of the system. Indeed, $\tau_{\mathrm{diss}}$ is of the same order of magnitude as the linear viscous dissipation by surface boundary layer of the main container mode $\tau_{\mathrm{diss}}=2\sqrt{2}/\left[k_{D/3}\sqrt{\nu\omega(k_{D/3}\mathrm{,}g^*)}\ \right]$ \cite{Lamb,DeikePRE12} ($\simeq 6$ s for 15$g$), whereas the nonlinear interactions are modified by this container mode (see below). $\tau_{\mathrm{diss}}$ and $\tau_{\mathrm{nl}}$ are also roughly independent of $g^*$ (see Supplemental Material). 


\paragraph*{3- and 4-wave interactions contributions.\textemdash}Assume a wave energy decay at the Fourier mode $f$ as
\begin{equation}
\frac{dE_f(t)}{dt}=-aE_f(t)-bE^2_f(t)-cE^3_f(t)
\label{decaymodel}
\end{equation}
with $a$, $b$ and $c$ positive values depending on $f$. The first term of the right-hand side corresponds to a usual viscous linear dissipation, the second and third terms modeling the nonlinear transfer from 3- and 4-wave interactions, respectively. Considering only one non-zero coefficient, $a$, $b$ or $c$, leads to usual decaying solutions: $E_f(t)/E_f(0)=\exp{[-t/\tau_{\mathrm{diss}}]}$, $\left[1+t/\tau^c_{nl}(f)\right]^{-1}$ and $\left[1+2t/\tau^g_{nl}(f)\right]^{-1/2}$, for only dissipation ($a \neq 0$), only 3-wave interactions ($b \neq 0$) and only 4-wave interactions ($c\neq 0$), respectively \cite{DeikeJFM15}. Since no power law in $t^{-1/2}$ or $t^{-1}$ is found experimentally in the fast decay, we have to take into account in Eq.\ (\ref{decaymodel}) the three non-zero coefficients at the same time. Fixing the viscous dissipation time $1/a$, independently of $f$ as found experimentally, and fitting each experimental Fourier mode by time-power laws give the nonlinear timescales for 3-wave interactions, $1/b$ ($\blacksquare$), and for 4-wave ones, $1/c$ ($\square$), versus $f$ as shown in Fig.\ \ref{fig05}.  In the capillary regime, we find that 4-wave interactions take place ($c\neq0$) and 3-wave interactions are negligible ($b \approx 0$), contrarily to weak turbulence predictions. In the gravity regime, we find that 3-wave interactions occur ($c\approx 0$, $b\neq 0$). Although forbidden by weak turbulence theory for plane gravity waves, 3-wave interactions are authorized theoretically in cylindrical containers where axisymmetric eigenmodes are important \cite{MichelPRF19}. Indeed, axisymmetric modes imply a new conserved quantity (angular pseudomomentum) \cite{MichelPRF19}. The large-scale axisymmetric modes being important in our study, they thus modify the type of interaction mechanism for gravity waves, and probably change the one for capillary waves. Such finite size effects on the wave interactions would deserve further studies with more accurate space-time measurements.


\paragraph*{Energy flux.\textemdash}One can finally estimate the mean cascading energy flux $\varepsilon$ from the wave energy decay \cite{DeikeJFM15}. Indeed, the wave energy (neglecting capillary waves) is $E(t)=g^*\sigma^2_{\eta}$, and the power budget (assuming no forcing and dissipation in the inertial range) then reads $dE(t)/dt=-\varepsilon$, quantities being per unit surface and fluid density. Combining the two expressions then leads to $\varepsilon=-g^*d\sigma_{\eta}^2(t)/dt|_{t=0}$. Thus, experimentally from the tangent at $t=0$, $\varepsilon$ is found to increase linearly with the apparent gravity $g^*$ as expected, and is found to be 2 orders of magnitude smaller than the critical flux breaking weak turbulence, $(\gamma g^*/\rho)^{3/4} \simeq 800{g^*}^{3/4}$ cm$^3$s$^{-3}$, regardless of the value of $g^*$ (see Supplemental Material).

\paragraph*{Conclusion.\textemdash}In this Letter, we reported the first observation of gravity-capillary wave turbulence in a high-gravity environment. This specific set-up favors the study of gravity wave turbulence in the laboratory by extending significantly its inertial range and the critical energy flux. We observe power-law wave spectra in both regimes independent of the gravity level. The timescale separation required by weak turbulence is verified experimentally by means of nonstationary experiments. We also show that large-scale container modes play an important role by notably modifying the type of nonlinear interactions. Tuning the gravity field appears as a promising solution to study in laboratory the large-scale properties (i.e., larger than the forcing scale) of gravity wave turbulence. More generally, identifying the different scenarios (inverse cascade, statistical equilibrium, condensate) governing large-scale properties of turbulent flows is of primary interest \cite{Alexakis18} in wave turbulence \cite{AnnenkovPRL06,DeikeEPL11,Michel17} and 3D turbulence \cite{Xia11,Dallas15,Falcon17}. Our approach could be applied in different fields involving wave turbulence with finite size effects such as Rossby inertial waves (limited by the finite diameter of the planets), long internal waves in the oceans, or plasma waves in tokamaks.

\begin{acknowledgments}
We thank European Space Agency (ESA) for the access to the LDC facility through the proposal CORA-GBF-2018. We thank M. M\'elard for his technical help on the experimental setup. SD is a FNRS Senior Research Associate. AD is a FNRS Research Fellow. Part of this work was supported by FSR PPOPFF C-15/17 (ULg), FWB and ``{\em Primes de soutien aux promoteurs de doctorat}'' (University of Li\`ege), the French National Research Agency (ANR DYSTURB project No. ANR-17-CE30-0004), and a grant of the Simons Foundation/MPS N$^{\rm o}$651463 (Wave Turbulence).
\end{acknowledgments}

\end{document}